\documentclass[12pt,aps,showpacs,eqsecnum,nofootinbib,floatfix]{revtex4}
\usepackage{bbding}
\usepackage{pifont}
\usepackage{subfigure}
\usepackage{epsfig}
\usepackage{CJK}
\usepackage{amsmath}
\usepackage{amsfonts}
\usepackage{amssymb}
\usepackage{color}
\usepackage{graphicx}
\usepackage{mathrsfs}

\def\be{\begin{equation}}
\def\ee{\end{equation}}
\def\ba{\begin{eqnarray}}
\def\ea{\end{eqnarray}}
\def\no{\nonumber}
\def\p{\partial}
\def\d{\frac}

\definecolor{dyellow}{rgb}{1.,0.8,.0}
\definecolor{myblue}{rgb}{.1,.1,.7}
\definecolor{dcyan}{rgb}{.0,.6,.6}
\definecolor{dmagenta}{rgb}{0.6,0.0,0.6}
\definecolor{brown}{rgb}{0.6,0.2,0.}
\definecolor{darkblue}{rgb}{.0,.0,0.5}
\definecolor{darkred}{rgb}{0.75,0.0,0.0}
\definecolor{orange}{rgb}{1.,.6,.0}
\definecolor{dorange}{rgb}{0.8,.4,.0}
\definecolor{darkgreen}{rgb}{0.0,0.6,0.0}
\definecolor{purple}{rgb}{.4,.0,.4}
\definecolor{lightgrey}{rgb}{0.7, 0.7, 0.7}
\definecolor{grey}{rgb}{0.4, 0.4, 0.4}

\def\d#1#2{\frac{\displaystyle #1}{\displaystyle #2}}

\newcommand\btd{\raise 2pt
\hbox{$\hat\bigtriangledown$}\hskip 1.5pt}
\newcommand\bt{\raise 2pt
\hbox{$\bigtriangledown$}\hskip 1.5pt}

\newcommand{\omits}[1]{}

\def\NPB{{Nucl. Phys.}~{\bf B}}
\def\PRD{{Phys. Rev.}~{\bf D}}

\def\CQG{{Class. Quant. Grav. }}

\def\JHEP{{JHEP}}

\begin{document}

\title{Continuous phase transition and critical behaviors of 3D black hole with torsion}

\author{Meng-Sen Ma}
\email{mengsenma@gmail.com}
\author{Fang Liu}
\author{Ren Zhao}
\email{zhao2969@sina.com}

\medskip

\affiliation{Department of Physics, Shanxi Datong
University,  Datong 037009, China\\
Institute of Theoretical Physics, Shanxi Datong
University, Datong 037009, China}

\begin{abstract}

we study the phase transition and the critical behavior of the BTZ
black hole with torsion obtained in $(1+2)$-dimensional Poincar\'{e}
gauge theory. According to Ehrenfest's classification, when the
parameters in the theory are arranged properly the BTZ black hole
with torsion may posses the second order phase transition which is
also a smaller mass/larger mass black hole phase transition.
Nevertheless, the critical behavior is different from the one in the
van der Waals liquid/gas system. We also calculated the critical
exponents of the relevant thermodynamic quantities, which are the
same as the ones obtained in the Ho\v{r}ava-Lifshitz black hole and
the Born-Infeld black hole.

\end{abstract}

\maketitle

\section{introduction}

The laws of black hole thermodynamics and the usual laws of
thermodynamics have very similar forms. The good agreement indicates
that black hole is also a thermodynamic system. Phase transitions
and critical phenomena are important characteristics of  usual
thermodynamics. Thus, the natural question to ask is whether there
also exists  phase transition in the black hole thermodynamics.
In fact, since the pioneering work of Davies\cite{Davies} and
  the well-known Hawking-Page phase transition\cite{hawking} were proposed, the question has been answered partly.
  The phase transitions and critical phenomena in \textit{four and higher} dimensional AdS black holes have been studied extensively\cite{Hut,Mazur,Lousto,Lemos}.
 Recently, some interesting works on asymptotically anti-de Sitter black holes have been done,
   which show that there exists phase transition similar to the van der Waals
liquid/gas phase
transition\cite{chamblin,chamblin1,Wu,Cvetic,Kastor,Dolan1,
RBM,Tian,RBM2}.

The properties of black hole are relevant to the dimension of
spacetime. Thus we want to know whether the similar phase transition
exists for lower dimensional black holes. The key advantage of lower
dimensional black holes lies in the simplicity of the construction.
Although just  mathematical abstraction, lower dimensional black
holes can be applied to physical reality in some special cases.
Hence it can be interesting to investigate the possibility that a
lower dimensional black hole  does exhibit a phase transition.
    BTZ black hole is an important solution of general relativity with negative cosmological constant in three-dimensional spacetime\cite{BTZ1,BTZ2}. The BTZ black hole is free of singularity and closely related with the recent developments
in gravity, gauge theory and string theory\cite{Witten}.

In general relativity and many other theories of gravity,
 curvature plays an essential role , while torsion has received
less attention. However, torsion  also has its geometrical meaning and plays some roles
in gravitation theory. Since the 1970s, many theories of gravity with torsion have been proposed, such as Poincar\'{e} gauge gravity, de Sitter gauge gravity,
teleparallel gravity, $f(T)$ gravity, etc. In particular,  Mielke and Baekler proposed a model of three-dimensional gravity with torsion ( the MB model),
 which also has  BTZ black hole as solution\cite{MB1,MB2,Hehl,Blagojevic1,Mielke}. This model aroused the following research on the thermodynamics of the BTZ black hole with torsion
  (\textit{BTZT black hole} for short) and
 AdS/CFT with torsion\cite{Blagojevic2,Blagojevic3,Blagojevic4}.

 In \cite{MMS} we have verified that for the BTZT black hole phase transition may exist. In this paper we
 will investigate the type of the phase transition and calculate the critical exponents. Although the BTZ solution is the same as the one
obtained in GR, the different actions will make their thermodynamics
very different. The modified action in the MB model will modify the
conserved charges such as mass and angular momentum. Correspondingly
the entropy of BTZ black hole and the first law of black hole
thermodynamics will also been changed.  It is shown that
 the heat capacity of the BTZ black hole in the MB model is not always positive any more, but changes signs at some points and may diverge at the critical point.
 Thus for the BTZT black hole in the MB model phase transition exists. According to Ehrenfest's classification we also consider the Gibbs free energy, the isothermal
 compressibility and the expansion coefficient as functions of temperature. It is shown that the kind of phase transition for the BTZT black hole belong to
 the second-order one or continuous one.

The paper is arranged as follows: in the next section we simply
introduce the MB model and its BTZ-like solution and the
corresponding thermodynamic quantities. In section 3 according to
Ehrenfest's classification we will analyze the type of the phase
transition of the BTZT black hole in the extended phase space. In
section 4 the case with non-extended phase space will be discussed
and the critical behaviors are investigated. We  make some
concluding remarks in section 5.

\section{Mielke-Baekler model and the 3D black hole with torsion}

First we should review the topological three-dimensional gravity model with torsion proposed by Mielke and Baekler\cite{MB1,MB2}, which is a natural
generalization of Riemannian GR with a cosmological constant.  Defining curvature and torsion 2-forms out of $\omega^a_{~b}$ and coframe $e^a$ by
\ba
 T^a&=&de^a+\omega^a_{~b}\wedge e^b, \\
 R^a_{~b}&=&d\omega^a_{~b}+\omega^a_{~c}\wedge \omega^c_{~b},
 \ea
the  gravitational action is written as
\be\label{action}
I=\int 2\chi e^{a}\wedge R_{a}-\frac{\Lambda }{3}\epsilon _{abc}e^{a}\wedge e^{b}\wedge e^{c}+\alpha
_{3}\left( \omega ^{a}\wedge d\omega _{a}+\frac{1}{3}\varepsilon _{abc}\omega
^{a}\wedge\omega ^{b}\wedge\omega ^{c}\right) +\alpha _{4}e^{a}\wedge T_{a}~,
\ee
where the dual expression, $R_a$ and $\omega_a$ are defined by $R^{ab}=\epsilon^{abc}R_c$ and $\omega^{ab}=\epsilon^{abc}\omega_c$.
In Eq.(\ref{action}) the first term corresponds to the Einstein-Cartan action, with $\chi=\frac{1}{16\pi G}$. The second one is the cosmological term. The last
two terms are the Chern-Simons term and the Nieh-Yan term, which should be given particular attention.

The Nieh-Yan(N-Y) form is a special 2-form only for the Riemann-Cartan geometry\cite{NY,Zanelli}.
On the four-dimensional manifold $M$ it can be written as
\ba
&&N=T^a\wedge T_a+R_{ab}\wedge e^a\wedge e^b=dQ_{NY}, \no\\
&&Q_{NY}=e^a\wedge T_a
\ea
The N-Y form is a kind of Chern-Simons form and will have its application
to manifolds with boundaries and reflect the role of torsion in geometry.

After variation to $\omega^a_{~b}$ and  $e^a$, two vacuum equations can be obtained from the MB action (\ref{action})
\ba
T^a&=&\frac{p}{2}\varepsilon^a_{~bc}e^b\wedge e^c, \\
R^a&=&\frac{q}{2}\varepsilon^a_{~bc}e^b\wedge e^c,
\ea
with the two constant coefficients $p, ~q$ defined by
$p=\frac{\alpha _{3}\Lambda +\alpha _{4}\chi}{\alpha
_{3}\alpha _{4}-\chi^{2}}$ and $q=-\frac{\alpha _{4}^{2}+\chi\Lambda }{\alpha _{3}\alpha
_{4}-\chi^{2}}$.

The curvatures in Einstein-Cartan geometry can be connected to their counterparts in Riemannian geometry.
In particular, in three-dimensional spacetime, the equations above can be simplified to  equations without torsion
\be\label{notorsion}
\tilde{R}^a=\frac{\Lambda_{eff}}{2}\varepsilon^a_{~bc}e^b\wedge e^c,
\ee
where $\tilde{R}^a$ is the  curvature without torsion and $\Lambda_{eff}=q-\frac{1}{4}%
p^{2}$ is the effective cosmological constant. One can let $\Lambda _{eff}=-\frac{1}{l^{2}}<0$ to construct an asymptotically anti-de Sitter space.

As in the three-dimensional Einstein equation, Eq.(\ref{notorsion}) has the well-known BTZ solution. But in this case, torsion is contained in the gravitational action.
 The metric is
\be
ds^{2}=-N\left( r\right) ^{2}dt^{2}+\frac{1}{N\left( r\right) ^{2}}%
dr^{2}+r^{2}\left( d\phi +N_{\phi }dt\right) ^{2}
\ee
where
\begin{equation}
N\left( r\right) ^{2}=\frac{r^{2}}{l^{2}}-M_0+\frac{J_0^{2}}{4r^{2}},\text{
\ }N_{\phi }\left( r\right) =\frac{J_0}{2r^{2}}~.
\end{equation}%
Here we have considered $8G=1$. This metric is the same as the one
in GR, except that  $l=1/\sqrt{-\Lambda _{eff}}$ here and a constant
torsion\cite{Hehl}. For this metric there are two horizons: the
outer one $r_+$ and the inner one $r_{-}$. From $N^2(r)=0$, one can
obtain the expressions of  both horizons: \be\label{horizon}
r^2_{\pm}=\frac{M_0l^2}{2}(1\pm \Delta), \quad
\Delta=[1-(J_0/M_0l)^2]^{1/2} \ee Conversely, $M_0$ and $J_0$ can be
expressed as follows: \be M_0=\frac{r_{+}^2+r_{-}^2}{l^2}, \quad
J_0=\frac{2r_+r_{-}}{l}. \ee

Hawking radiation is just a kinematic effect, which only
depends on the event horizon and is irrelevant to the dynamical
equations and the gravitational theories. Therefore the temperature
of BTZ black hole in the MB model has the similar form as in GR,
which is
\be \label{temp}
T=\frac{r_{+}^2-r_{-}^2}{2\pi l^2r_{+}}
\ee
Certainly because of the existence of $l$, the temperature is relevant to the coefficients $\alpha_3,\alpha_4, \Lambda$ of MB Lagrangian.
Define
\be
\Omega_H=-\frac{g_{t\phi}}{g_{\phi\phi}}|_{r_{+}}=\frac{J_0}{2r_+^2}
\ee
which can be regarded as the angular velocity of BTZ black hole.

Because of the existence of the topological terms, the asymptotically behavior is different from the one for Einstein-Cartan theory.
 Blagojevic et.al have proved that the gravitational conserved charges in the MB model should be\cite{Blagojevic2,Blagojevic3,Blagojevic5}
\be\label{MJ}
M=M_0+2\pi\alpha_3(\frac{pM_0}{2}-\frac{J_0}{l^2})=a
M_0-\d{b}{l^2}J_0, \quad J=J_0+2\pi\alpha_3(\frac{p J_0}{2}-M_0)=a
J_0-b M_0
\ee where we have defined $a=1+ \pi\alpha_3p,
~b=2\pi\alpha_3$. Obviously when $\alpha_3=0$ they will return to
their conventional interpretation as energy and angular momentum, as
with the BTZ metric in general relativity.

Correspondingly the entropy can be derived
\be\label{entropy}
S=4\pi
r_{+}+4\pi^2\alpha_3(pr_{+}-\frac{2r_{-}}{l})=4\pi(a
r_{+}-\d{b}{l}r_{-})
\ee
It differs from the Bekenstein-Hawking
result by an additional term and will coincides with Solodukhin's
result if $p=0$\cite{SN}. Black hole entropy is not always equated
with one quarter of the event horizon area. In fact it is related to
the gravitational theory under consideration.  It can be easily
verified  that in the MB model the entropy, temperature, and the
conserved charges not only satisfy the first law of thermodynamics
\be dM=TdS+\Omega_{H} dJ \ee
but also fulfill the Smarr-like formula
\be
M=\frac{1}{2}TS+\Omega_{H}J \ee
This further implicates that with
torsion the BTZ black hole can still be treated as a thermodynamic
system and the thermodynamic laws still hold. It should be noted
that in the expression of the entropy of the BTZ black hole  no torsion
exists explicitly, only $\alpha_4$ in $p$ implicitly. In particular,
when $\alpha_3=0$ the entropy in Eq.(\ref{entropy}) returns to the
usual BTZ black hole entropy. It means that the N-Y term $\alpha
_{4}e^{a}\wedge T_{a}$  influences the conserved charges and the
exact form of entropy only when the CS term exists.

\section{phase transition in extended phase space}

Ehrenfest had ever attempted to classify the phase transitions . Phase transitions connected with
  an entropy discontinuity are called discontinuous or first order phase transitions, and phase transitions where the entropy is continuous are called continuous or second/higher order phase transitions. More precisely, for the first-order phase transition the Gibbs free energy $G(T,P,...)$ should be continuous and
  its first derivative with respect to the external fields:
  \be
  S=-\left.{\frac{\partial G}{\partial T}}\right|_{(P,...)}, \quad  V=\left.{\frac{\partial G}{\partial P}}\right|_{(T,...)}
  \ee
  are discontinuous at the phase transition points.

 For the second-order phase transition the Gibbs free energy $G(T,P,...)$ and its first derivative are both continuous,
 but the second derivative of $G$ will diverge at the phase transition points like the specific heat $C_P$,
 the compressibility $\kappa$, the expansion coefficient $\alpha$:
 \be
 C_P=T\left.{\frac{\partial S}{\partial T}}\right|_P=-T\left.{\frac{\partial^2 G}{\partial T^2}}\right|_P,
 \kappa =-\frac{1}{V}\left.{\frac{\partial V}{\partial P}}\right|_T=-\frac{1}{V}\left.{\frac{\partial^2 G}{\partial P^2}}\right|_T ,
 \alpha =-\frac{1}{V}\left. {\frac{\partial V}{\partial T}} \right|_P=-\frac{1}{V}\frac{\partial^2 G}{\partial P\partial T}
 \ee
In this sense, because the heat capacity is always positive, there
is no second order phase transition for BTZ black hole obtained in
GR. This property of BTZ black hole can also be verified by the
method of thermodynamic curvature\cite{Cai,Quevedo}.

To utilize Ehrenfest's classification, we consider variable cosmological constant and relate it to the pressure\cite{Kastor,Dolan1,
RBM,RBM2,Dolan2,Dolan3}. The first law of thermodynamics for the BTZT black hole should be
\be\label{flcosm}
dM=TdS+\Omega dJ+V dP
\ee
where $P=\d{1}{8\pi l^2}=-\d{\Lambda_{eff}}{8\pi}$,  and $V=\left.{\d{\p M}{\p P}}\right|_{S,J}$ is the corresponding thermodynamic volume. Therefore the mass of black hole is no more internal energy, but should be interpreted as the thermodynamic
enthalpy, namely $H=M(S,P,J)$\cite{Kastor,Dolan1,Dolan2,Dolan3}. The first law of black hole thermodynamics represented by the internal energy $U(S,V,J)$ reads
\be\label{flenergy}
dU=TdS+\Omega dJ-PdV
\ee
where $U=H-PV$.

For BTZT black hole one can  express mass $M$ as functions of
$S, J, P$, which are
\be\label{m}
H_{\pm}=M_{\pm}=\frac{1}{8\pi^2b^2}\left[{a S^2+8\pi^2ab J }
\pm  {S\sqrt{(a^2-8\pi b^2P)(S^2+16\pi^2b J)}}\right]
\ee
One can  substitute the expressions of $S,M,J$ into Eq.(\ref{m}) to
test and verify it directly. This result can also be verified easily
by differentiating the mass $M_{\pm}$ with the entropy $S$ to get
the Hawking temperature, Eq.(\ref{temp}).

It should be noted that when $M$ is expressed with $r_{+}, r_{-},l$, its form is unique. When we express it with the thermodynamic quantities,
two different forms $M_{\pm}$ appear. In fact $M_{\pm}$ depend on the relation between $al$ and $b$. They are established under different conditions:
\begin{itemize}
  \item $b^2 \leq a^2l^2$ or $b^2 \leq \d{a^2}{8\pi P}$, the expression $M_{-}$ is right. At this time, to keep the expression in square root have
  physical meaning, $ S^2+16\pi^2b J \geq 0$ should also be satisfied.

  \item $b^2 \geq a^2l^2$ or $b^2 \geq \d{a^2}{8\pi P}$, the expression $M_{+}$ should be used. Similarly, to keep the expression in square root have
  physical meaning, $ S^2+16\pi^2b J \leq 0$ should  be satisfied.

 \end{itemize}
According to Eqs.(\ref{MJ}),(\ref{entropy}),when $|b|\geq |a|l$, $M,
J, S$ may be negative. In fact for gravities with higher derivative
terms there is the possibility for negative entropy and energy which
depend on the parameters of higher derivative terms\cite{Nojiri}.
Although black holes behave as thermodynamic systems, they also show
some exotic behaviors, the most known one is the entropy of black
holes is proportional to area and not the volume. Therefore it is
understandable if black holes exhibit some strange thermodynamic
properties.  Below we will show that the condition $|b|\geq |a|l$ is
the key for the BTZT black hole to have phase transition.

The temperature of
BTZT black hole can be evaluated according to $S,P,J$:
\be
\label{tem} T_{-}(S,J,P)=\left.\d{\p M_{-}}{\p S}\right|_{P,J} \text{or} \quad T_{+}(S,J,P)=\left.\d{\p M_{+}}{\p S}\right|_{P,J}
\ee

They look different when expressed with the thermodynamic quantities $S,P,J$ because they correspond to different conditions for the parameters
$a,b,l$. When replacing the thermodynamic variables $(S,J,P)$ with
the geometric ones $(r_+, r_{-},l)$, the two expressions can be
unified and Eq.(\ref{temp}) can turn up.

According to Eqs.(\ref{m}),(\ref{tem}), the specific heats at constant pressure and constant angular momentum can be calculated easily:
\be
C_{-}=\left.\d{\p M_{-}}{\p T{-}}\right|_{P,J}=T_{-}\frac{1}{\left.{\frac{\partial T_{-}}{\partial S}}\right|_{P,J}},
\quad
C_{+}=\left.\d{\p M_{+}}{\p T{+}}\right|_{P,J}=T_{+}\frac{1}{\left.{\frac{\partial T_{+}}{\partial S}}\right|_{P,J}}
\ee
With the geometric quantities, the heat capacity can be written as
\be\label{hc}
C_P=C_{\pm}=\frac{4 \pi  r_{+}^2 \left(r_{+}^2-r_{-}^2\right) \left(b^2-a^2 l^2\right)}{l \left(b~ r_{-} \left(3 r_{+}^2+r_{-}^2\right)
-a l r_{+} \left(r_{+}^2+3 r_{-}^2\right)\right)}
\ee

It is more appropriate to study the phase transition of the BTZT black holes according to geometric quantities $r_{+},~r_{-}$. Because the conditions $r_{+}\geq r_{-}, ~M_0>0, ~J_0>0$ must be fulfilled.
If employing the thermodynamic quantities completely, one may omits these conditions. \omits{In fact there exists inconsistency indeed.
We will show that in the Appendix.} Below we will set $l=1$ and analyze the phase transition of the BTZT black hole numerically.
According to Ehrenfest's classification, we should first derive the Gibbs free energy $G$:
\be
G=H-TS=M-TS
\ee

\subsection{$a^2l^2\geq b^2$}
\begin{figure}[!htbp]
\centering
\includegraphics[angle=0,width=5cm,keepaspectratio]{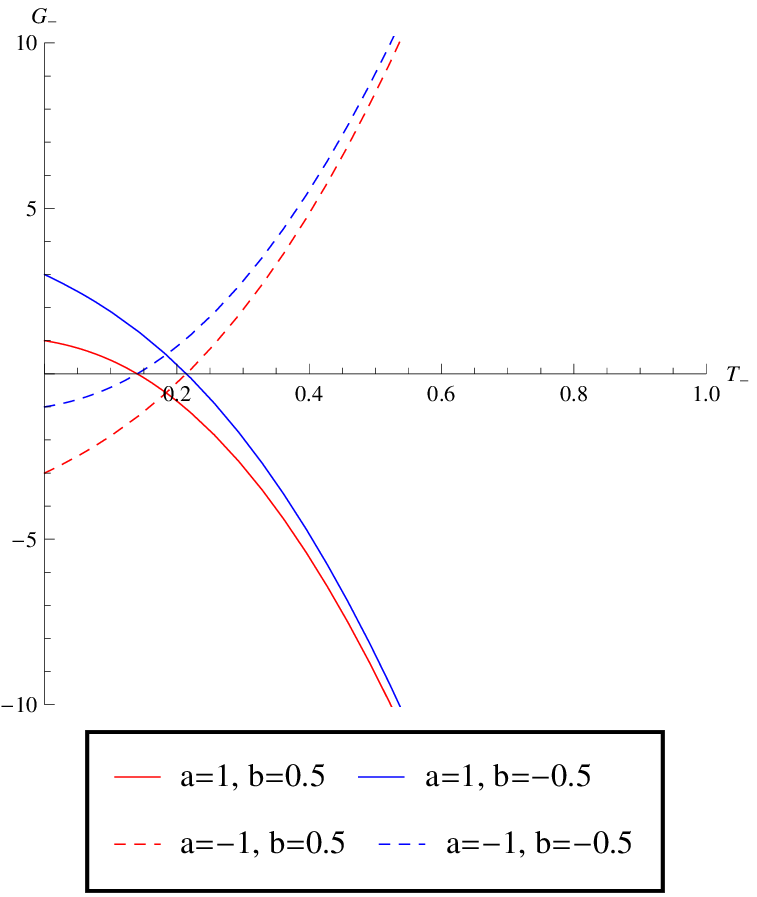}
\includegraphics[angle=0,width=5cm,keepaspectratio]{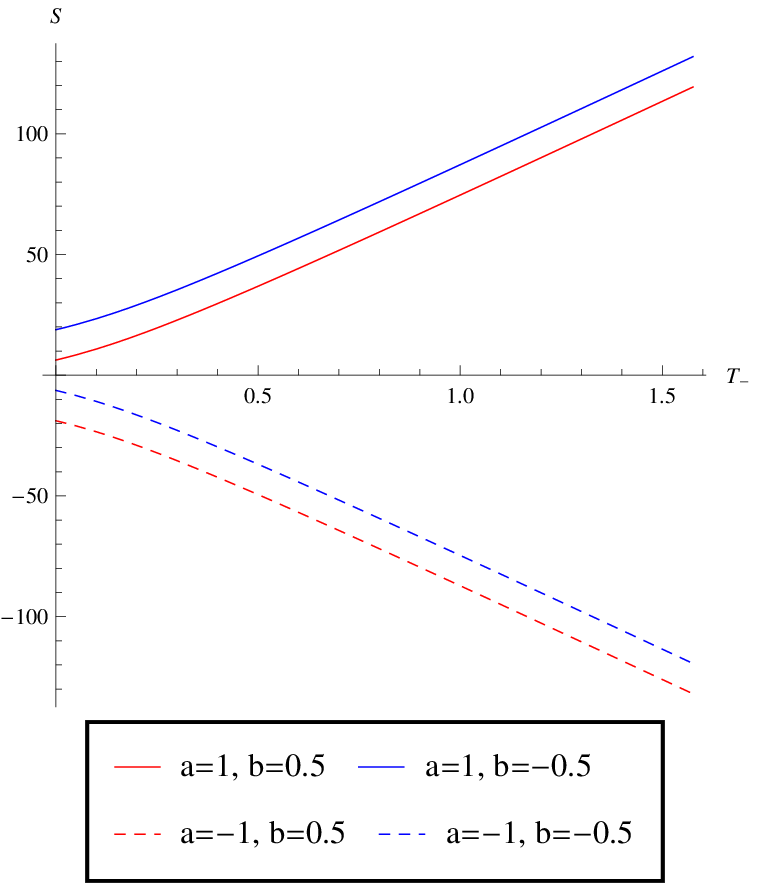}
\includegraphics[angle=0,width=5cm,keepaspectratio]{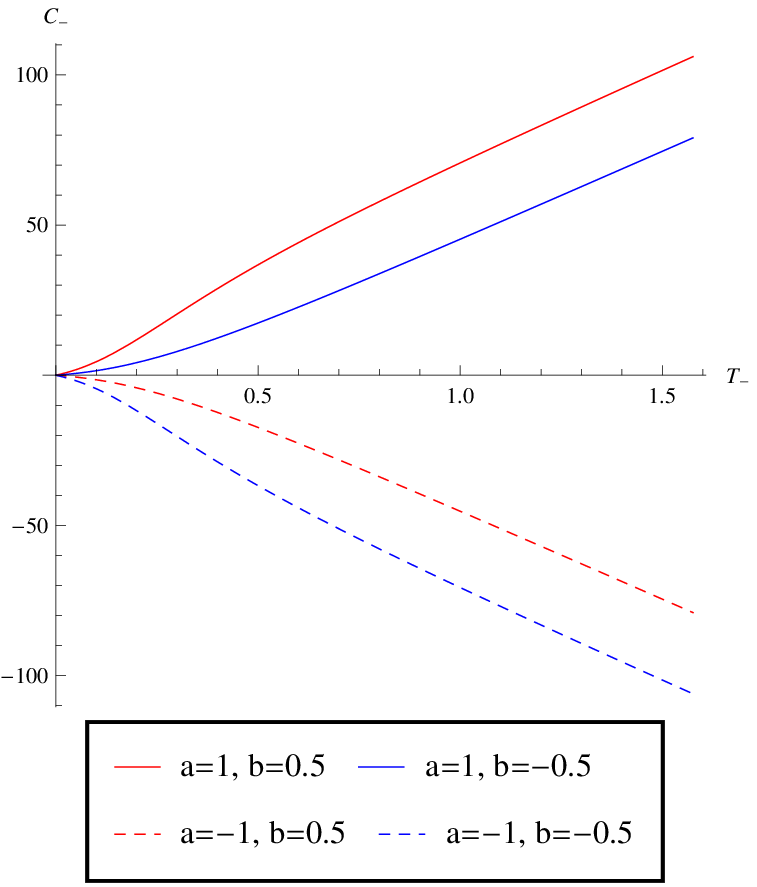}
\caption[]{\it The Gibbs free energy, entropy and heat capacity at constant pressure as functions of temperature for BTZT
 black hole for the choices of $ ~l=1, ~r_{-} =1,~a=\pm 1,~b=\pm 0.5 $ and $r_{+}\geq r_{-}$. For $C_{-}$ no divergent point exists. }
\label{GSCN}
\end{figure}

One can easily plot the $G_{-}-T_{},~S-T_{},~ C_{-}-T_{}$ curves
  as shown in Fig.\ref{GSCN}.  Obviously $G_{-},~ S,~ C_{-}$ are all continuous function of
   temperature $T$ . No turning point and divergence turn up,
    which means no second-order phase transition happens in this case. This conclusion can also
     be supported by the method of thermodynamic curvature\cite{Quevedo,Sengupta}.

\subsection{$a^2l^2\leq b^2$}

\begin{figure}[!htbp]
\center{\subfigure[~$ a=1, ~b=-2 $] {
\includegraphics[angle=0,width=6cm,keepaspectratio]{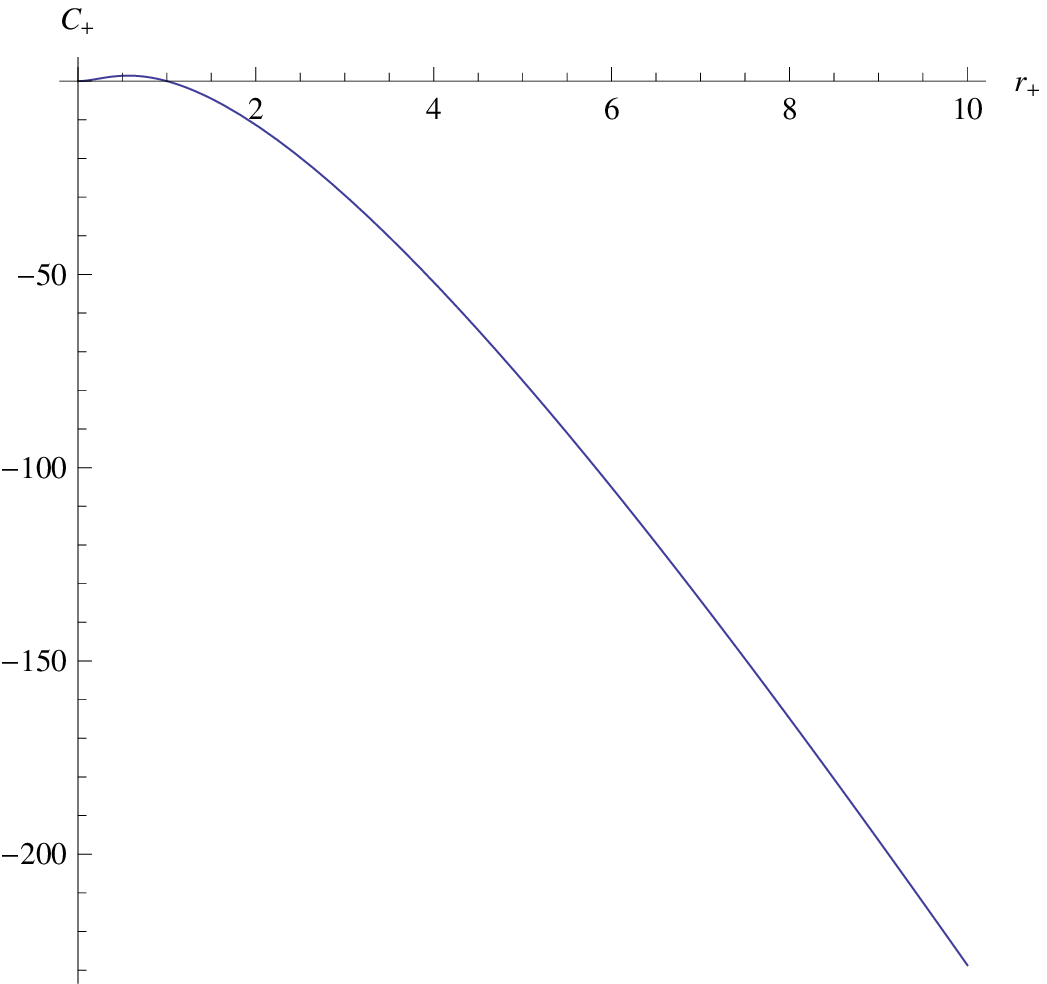}}\hfill
\subfigure[~$ a=-1, ~b=2 $] {
\includegraphics[angle=0,width=6cm,keepaspectratio]{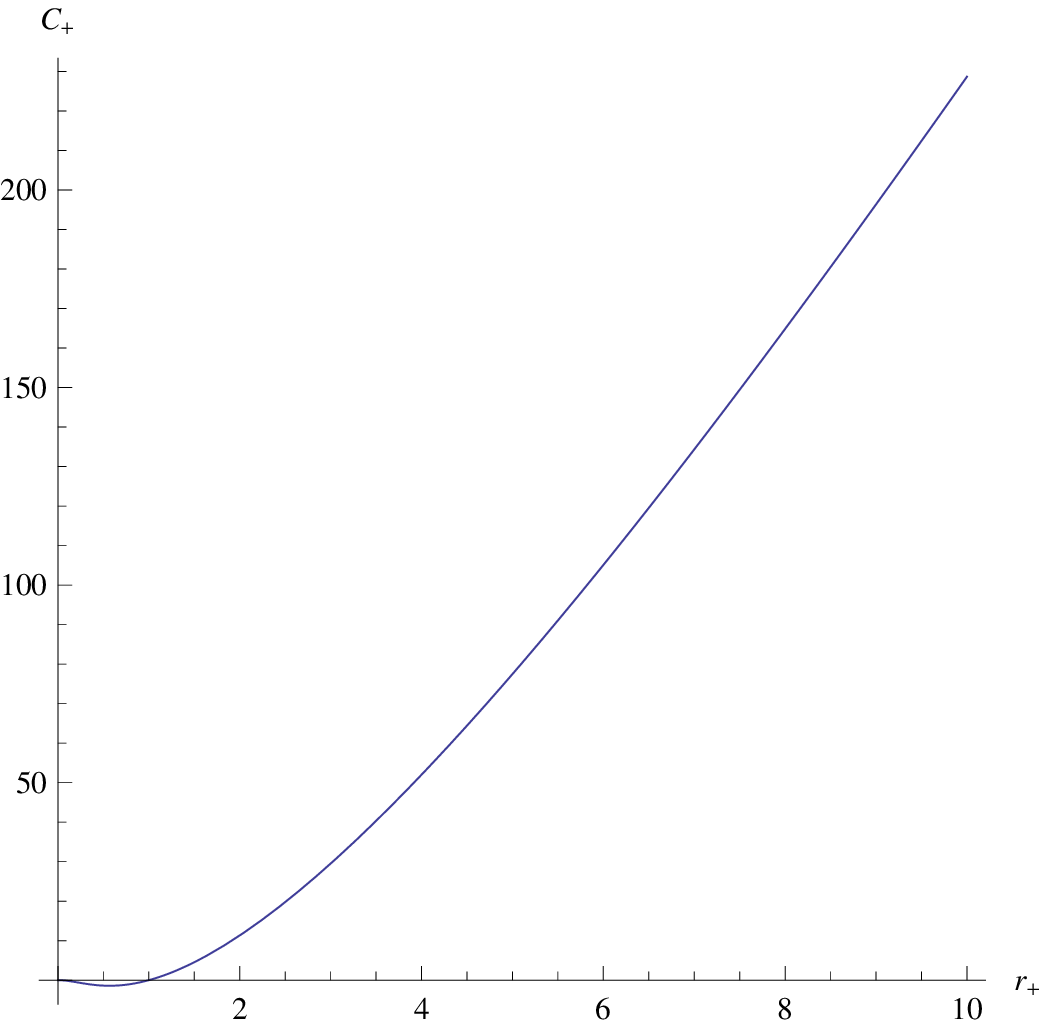}}\\
\subfigure[~$ a=1, ~b=2 $] {
\includegraphics[angle=0,width=6cm,keepaspectratio]{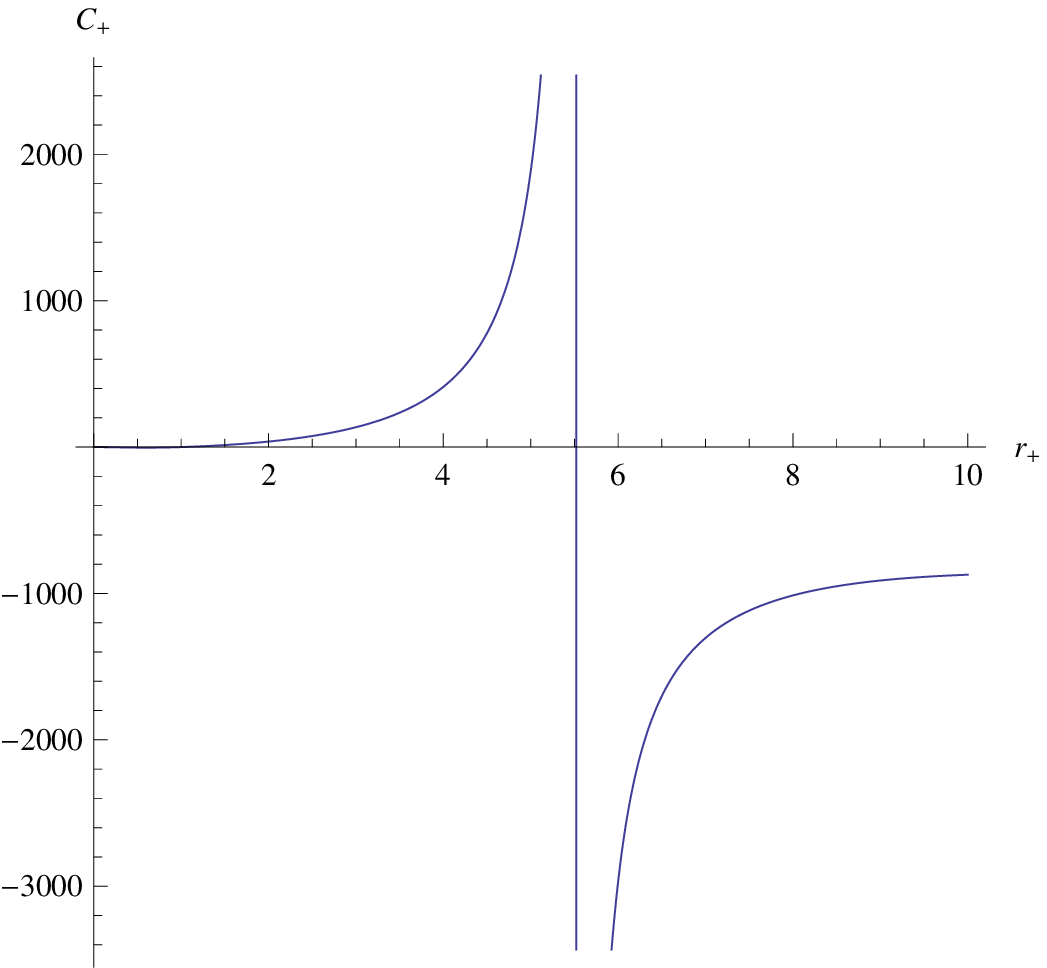}}\hfill
\subfigure[~$ a=-1, ~b=-2 $] {
\includegraphics[angle=0,width=6cm,keepaspectratio]{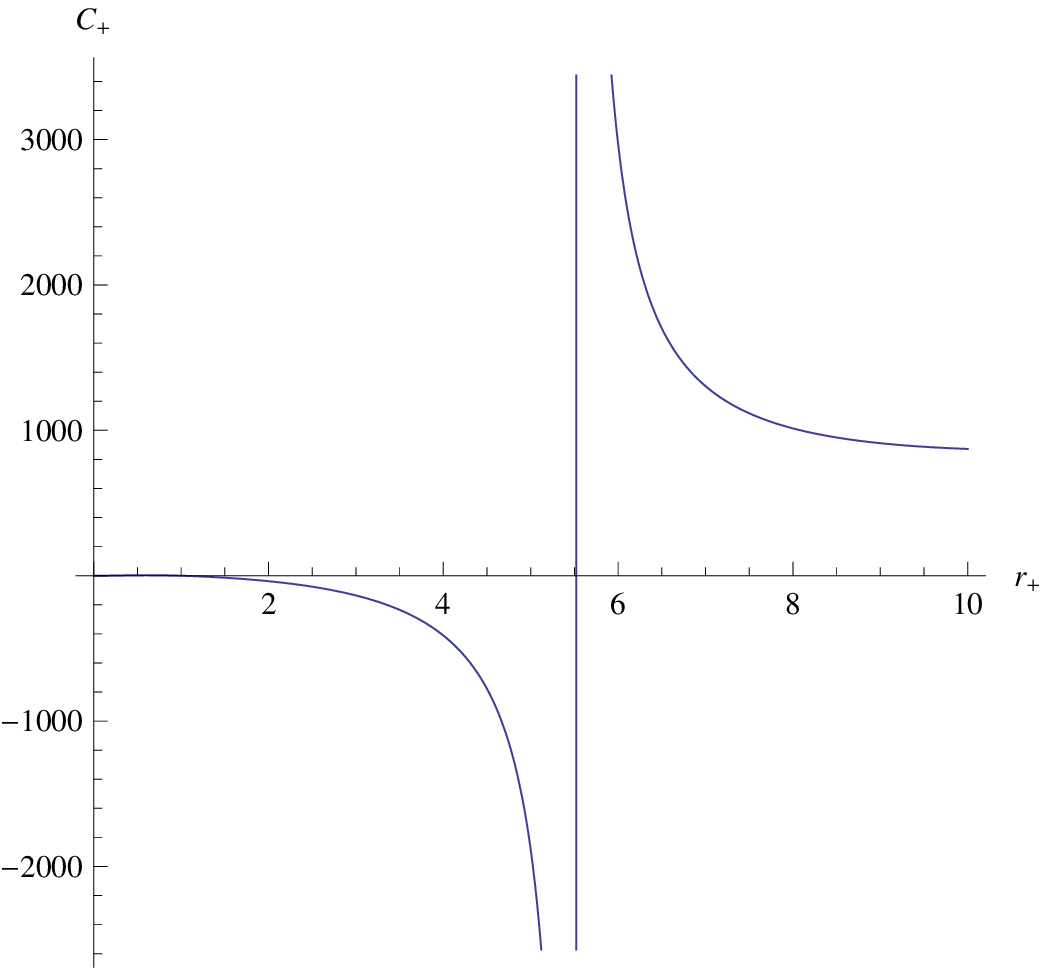}}
\caption[]{\it The heat capacity at constant pressure and angular
momentum as functions of $r_{+}$ for BTZT black hole for the choices
of $ ~l=1,~r_{-}=1, ~a=\pm 1,~b=\pm 2 $. In the cases $(c),(d)$
there exist the divergent points for $C_{+}$. } \label{CPR}}
\end{figure}

In this case we first plot the $C_{+}-r_{+}$ curves for different values of $a,~b$. In Fig.\ref{CPR}(a) and Fig.\ref{CPR}(b) we show that there is no divergent point when $a,~b$ take
opposite signs. The phase transition may happen only when $a,~b$ are both positive or negative. Under the given conditions one can easily derive
 the position of the divergent point, $r_{c}\approx 5.522$. Obviously the phase transition is  a smaller mass/larger mass black hole phase transition.
 When $a>0,~b>0$, the smaller black hole is stable because  the heat capacity is positive, when $a<0,~b<0$,  the other way around, the larger black hole is stable.

\begin{figure}[!htbp]
\center{\subfigure[~$ a=1, ~b=2 $] {
\includegraphics[angle=0,width=7cm,keepaspectratio]{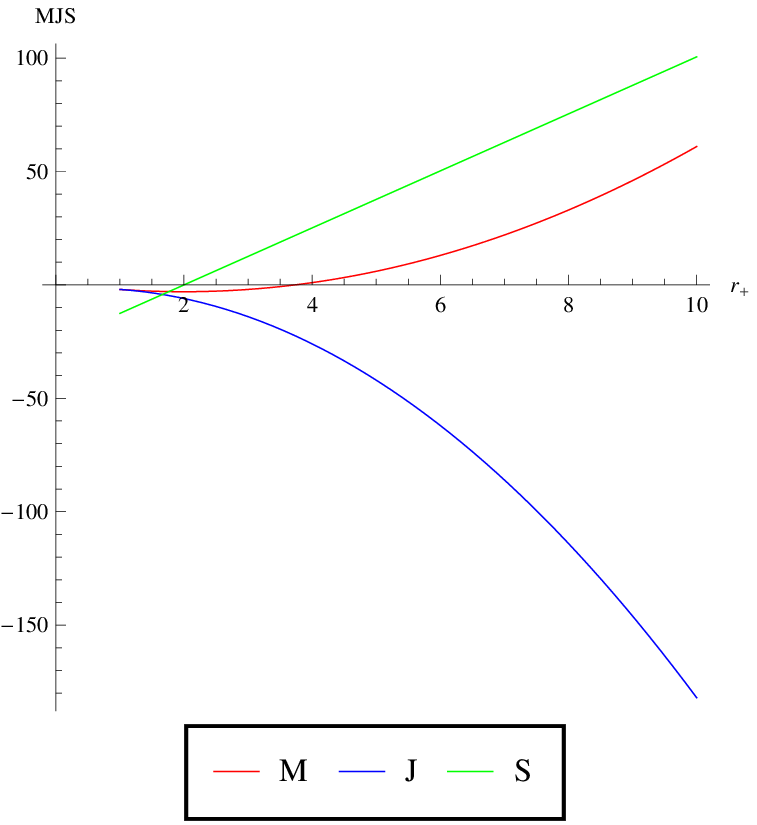}}\hfill
\subfigure[~$ a=-1, ~b=-2 $] {
\includegraphics[angle=0,width=7cm,keepaspectratio]{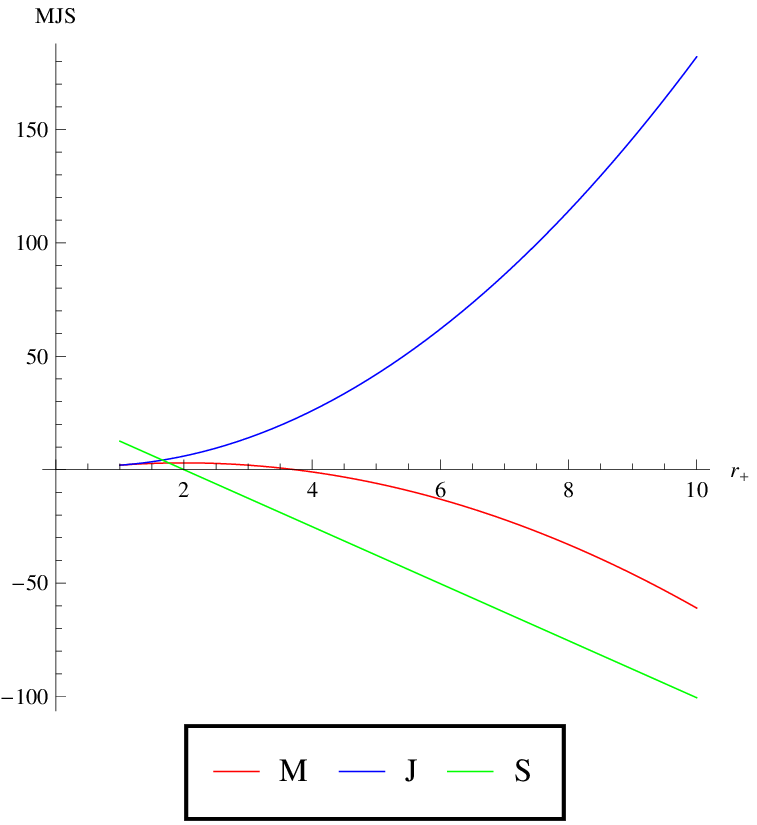}}
\caption[]{\it The conserved charges $M,~J$,and the entropy  as functions of $r_{+}$ for BTZT
 black hole for the choices of $ ~l=1, ~r_{-} =1,~a=\pm 1,~b=\pm 2 $ and $r_{+}\geq r_{-}$.  }\label{MJS}}
\end{figure}

One should further analyze the value of $M, ~J,~S$. We only consider the two cases which have the phase transitions. In Fig.\ref{MJS} it is shown that
under the condition $a^2l^2\leq b^2$ when $a,~b$ are both positive, the corrected mass $M$ and the entropy $S$ are positive at the divergent point
$r_{c}$, while  $J$ is negative; for the case with negative $a,~b$, the situation is just the opposite. Thus we conclude that for the BTZT black hole
if the phase transition can happen, the $M,~J,~S$ cannot be all positive.

\begin{figure}[!htbp]
\centering
\includegraphics[angle=0,width=7cm,keepaspectratio]{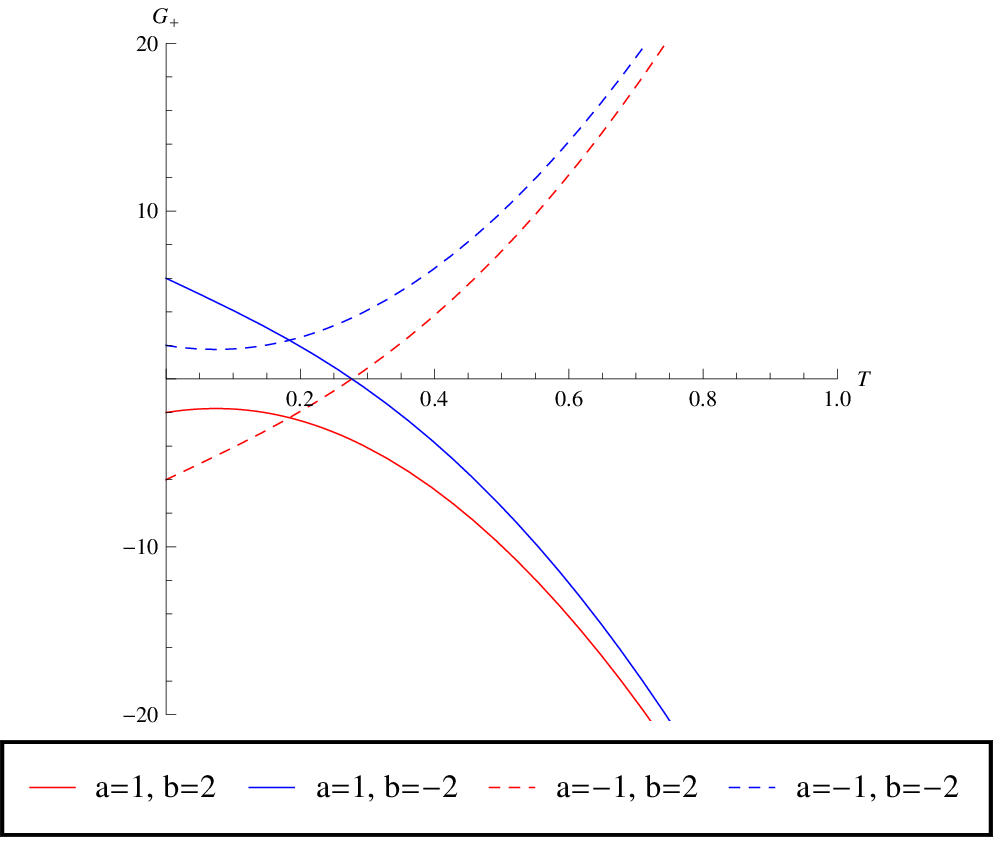}\hfill
\includegraphics[angle=0,width=7cm,keepaspectratio]{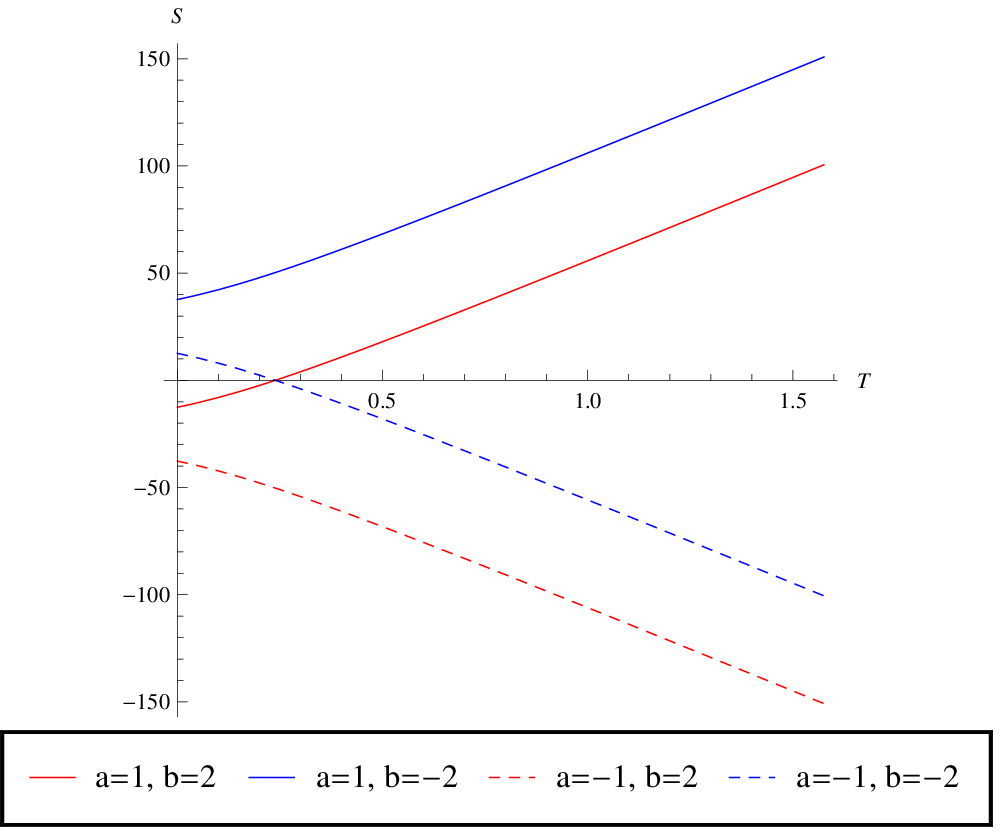}\\
\includegraphics[angle=0,width=8cm,keepaspectratio]{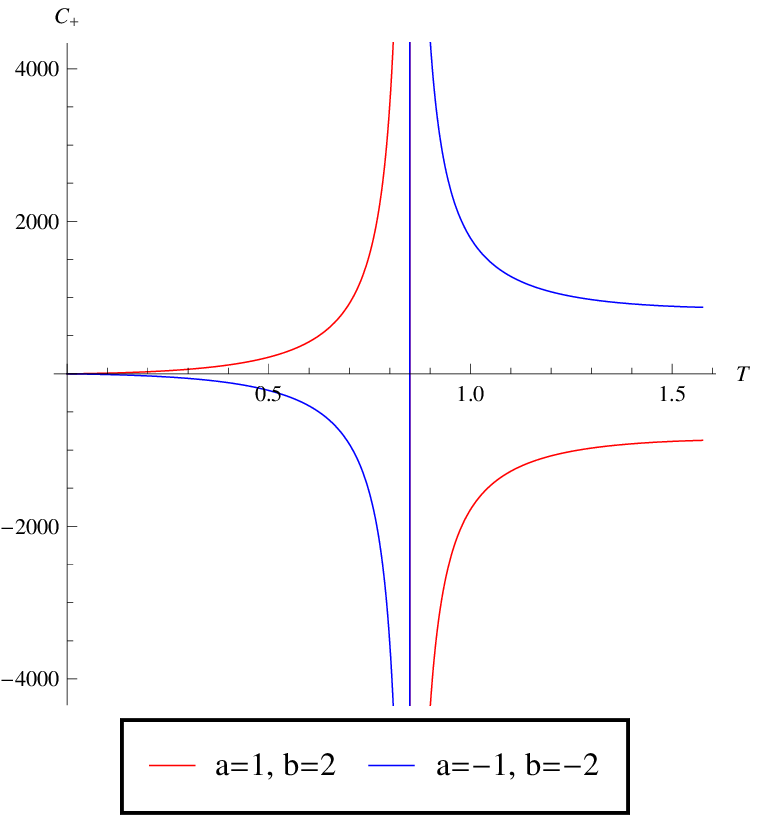}\\
\includegraphics[angle=0,width=7cm,keepaspectratio]{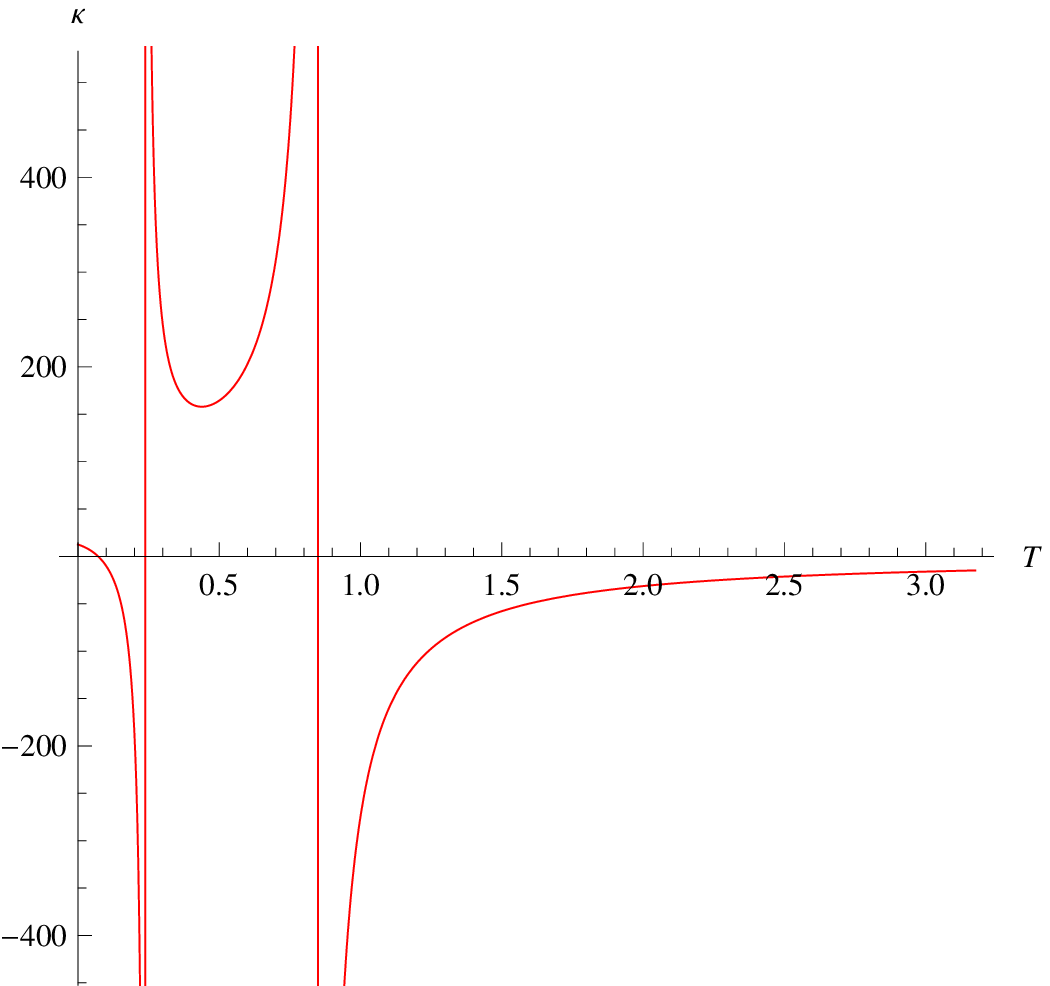}\hfill
\includegraphics[angle=0,width=7cm,keepaspectratio]{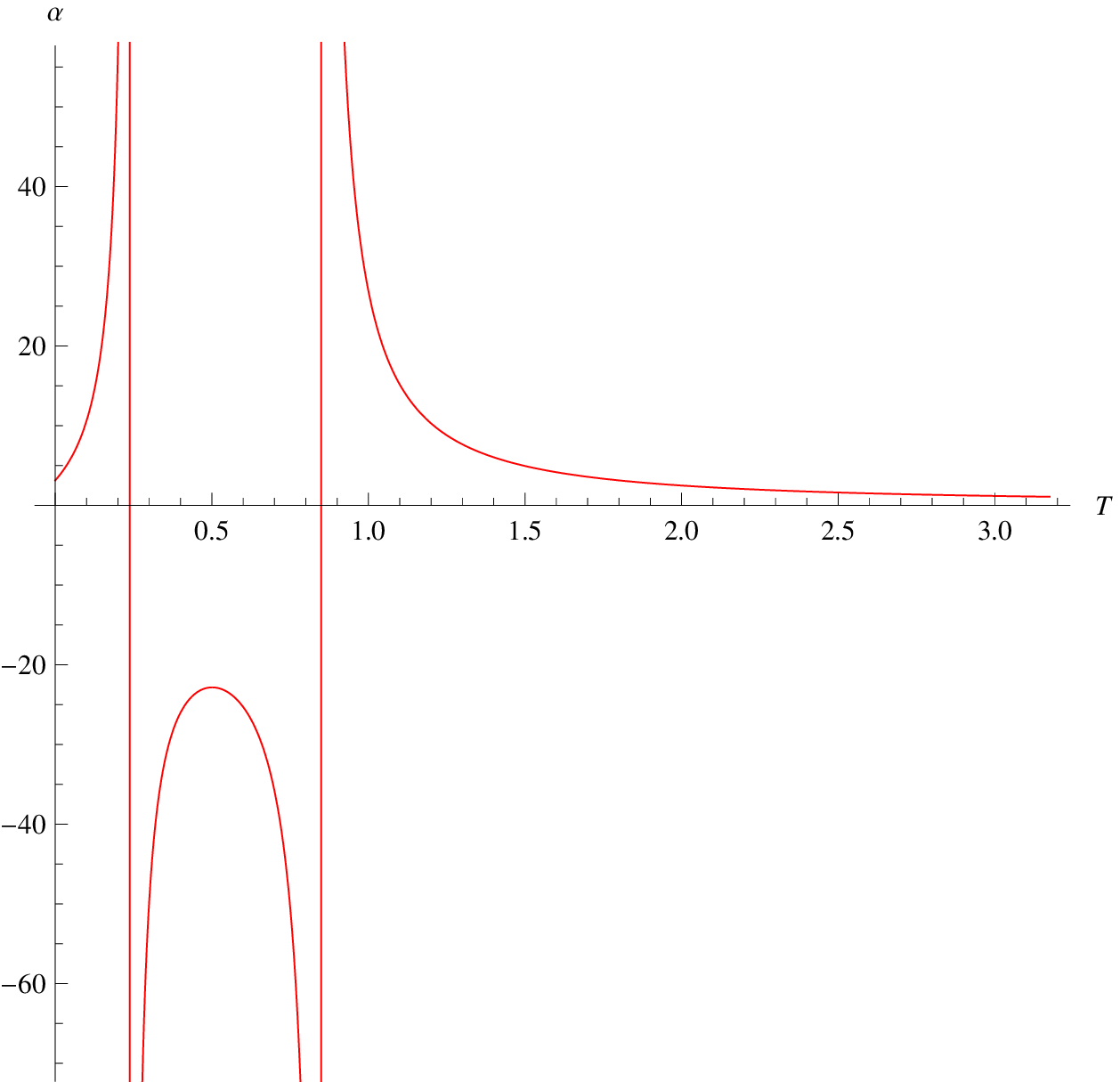}
\caption[]{\it The Gibbs free energy, entropy, the heat capacity at constant pressure and the isothermal compressibility and the expansion coefficient
 as functions of temperature for BTZT
 black hole for the choices of $ ~l=1, ~r_{-} =1,~a=\pm 1,~b=\pm 2 $ and $r_{+}\geq r_{-}$. For the two cases $a=1,~b=2$ and $a=-1,~b=-2$,
 the $\kappa-T$ and the $\alpha-T$ curves are both concurrent respectively.}
\label{GSCP}
\end{figure}

Now we investigate the type of the phase transition for the BTZT black hole according to Ehrenfest's classification. To calculate the isothermal
 compressibility  $\kappa$ and the expansion coefficient $\alpha$ we should first obtain the thermodynamic volume.
 \be V_{+}=\left.{\frac{\partial M_{+}}{\partial
P}}\right|_{S,J},\quad V_{-}=\left.{\frac{\partial M_{-}}{\partial
P}}\right|_{S,J}
\ee
Although $M_{+}$ and $M_{-}$ are different,
when considering the conditions for $M_{\pm}$ one can find that
\ba\label{vol}
V=V_{+}=V_{-}&=&\frac{S }{2 \pi  }\sqrt{\frac{16\pi ^2 b J
+S^2}{a^2-8\pi b^2 P  }}
\ea
Inversely, one can derive the pressure
\be\label{pre}
P=\frac{-16 \pi ^2 b  J S^2-S^4+4 \pi ^2 a^2 V^2}{32 \pi ^3 b^2
V^2}
\ee
which must be greater than zero. Thus $ 4 \pi^2 a^2 V^2-16
\pi^2 b J S^2-S^4 \geq 0 $ should be satisfied.
Because $V=V(S,P,J)$,
\be
\left.{\frac{\partial V}{\partial P}}\right|_T=\left.{\frac{\partial V}{\partial P}}\right|_S+\frac{\partial V}{\partial S}\left.{\frac{\partial S}{\partial P}}\right|_T
\ee
and
\be
\left.{\frac{\partial V}{\partial T}}\right|_P=\frac{\partial V}{\partial S}\left.{\frac{\partial S}{\partial T}}\right|_P
\ee
According to the above equations one can derive the isothermal compressibility $\kappa$ and the expansion coefficient $\alpha$.
We can plot the curves of $G-T,~S-T,~C-T,~\kappa-T,~\alpha-T$ in Fig.\ref{GSCP}. Similarly, only when the parameters $a,~b$ have the same sign, the phase transition for the BTZT black hole can turn up. The critical temperature lies at $T_{c}=0.85$. As is shown in the figure, the Gibbs free energy and the entropy are continuous functions of temperature. While the heat capacity, the isothermal compressibility and the expansion coefficient all diverge at the critical point. Therefore the phase transition at this critical point is the second-order phase transition or continuous one.

According to Eqs.(\ref{vol}),~(\ref{pre}) and $U=H-PV$, we can obtain the internal energy
\be
U=U(S,V,J)=-\frac{\left(S^2-2 \pi  a V\right) \left(-2 \pi  a V+16 \pi ^2 b J+S^2\right)}{32 \pi ^3 b^2 V}
\ee
From which one can easily derive the temperature as functions of $S,~V,~J$:
\be\label{tempv}
T(S,V,J)=\d{S (2 a \pi V-8\pi^2 b J  - S^2)}{8 b^2 \pi^3 V}
\ee
From  Eq.(\ref{pre}) and Eq.(\ref{tempv}), one can derive the equation of state between the pressure $P$, the temperature $T$ and the volume $V$ by eliminating $S$. Thus one can obtain the pressure $P$ as function of $V,T,J$ in principle. But the expression is too lengthy and obviously it will depend on the value of $J$. Below we will analyze the $P-V$ relation by means of the static scaling law\cite{Stanley}. Dimensional analysis implies that the $P$ and $T$ are both homogeneous functions of the variables $S, V, J$, since $P\rightarrow P$, $T\rightarrow \lambda T$ when $V \rightarrow \lambda^2 V, S \rightarrow \lambda S, J \rightarrow \lambda^2 J$. Thus $P$ and $T$ are in fact the functions of two independent variables.
The same logic also applies to the internal energy $U$. So we can take advantage of the scaling character to redefine the functions and the variables. One can take
\be
t=\frac{T}{S},\quad  p=P, \quad v=\frac{V}{S^2}, \quad j=\frac{J}{S^2}
\ee
In this way the entropy $S$ can be eliminated in Eqs.(\ref{pre}), (\ref{tempv}) and they are simplified to be
\ba
t&=&\frac{2\pi v-8 \pi^2\beta j -1}{8 \pi^3\beta^2 v}\no\\
p&=&-\frac{16 \pi^2\beta j +1-4 \pi^2 v^2}{32 \pi^3\beta^2  v^2}
\ea
Further removing the $j$ and combing the two equations together, one can get
\be\label{pv}
p=\frac{t}{2 v}+\frac{1}{8 b^2 \pi ^2}-\frac{1}{8 b^2 \pi ^2 v}+\frac{1}{32 b^2 \pi ^3 v^2}
\ee
The great advantage of the above relation lies at it is irrelevant to $J$ and is more universal. Although the $p-v$ structure is not the same as
the $P-V$ structure, it can reflect some properties of the system. The critical point occurs at the point where
\be
\frac{\p p}{\p v}=0, \quad \frac{\p^2 p}{\p v^2}=0
\ee
But unfortunately, the above equations do not have solutions. One can also easily plot the $p-v$ curves of Eq.(\ref{pv})
 at different temperatures. We can draw conclusions that  for the BTZT black hole there is no the similar phase structure and critical behavior
to the van der Waals liquid-gas system.

\section{critical behaviors in non-extended phase space}
In the non-extended phase space, the $l$ or $\Lambda$ should be considered as constant. Thus Eq.(\ref{m}) is modified to
\ba\label{m1}
M_{\pm}=\frac{1}{8\pi^2\beta^2}\left[{a S^2+8\pi^2ab J }
\pm  {\d{S}{l}\sqrt{(a^2l^2-b^2)(S^2+16\pi^2b J)}}\right]
\ea
When considering the $J-\Omega$ relations some rotating black holes such as Kerr-AdS black hole will exhibit similar critical behavior to the
van der Waals liquid-gas system\cite{Banerjee,WXN}. Thus we analyze  for the BTZT black hole whether there are similar conclusion.  The $J-\Omega$ relation
can be easily obtained:

\be\label{JW}
J=-\frac{4 \pi ^2 T^2 \left(-2 a l^2 \Omega +b l^2 \Omega
^2+b\right)}{\left(l^2 \Omega ^2-1\right)^2} \ee
One may try to derive the critical point according to
\be
\left.{\frac{\partial J}{\partial \Omega}}\right|_T=0, \quad \left.{\frac{\partial^2 J}{\partial \Omega^2}}\right|_T=0
\ee
But it can be easily verified that also no solution exists. In this case the heat capacity $C$ is the same as Eq.(\ref{hc}).
However, the isothermal compressibility should be defined as\cite{WXN}
\be
\kappa_{T}=\left.{\frac{\partial \Omega}{\partial J}}\right|_T=-\frac{\left(l^2 \Omega ^2-1\right)^3}
{8 \pi ^2 l^2 T^2 \left(3 a l^2 \Omega ^2+a-b l^2 \Omega ^3-3 b \Omega \right)}
\ee
We plot the $\kappa_{T}-T$ curves as in Fig.5. In this case the critical temperature still lies at $T_{c}=0.85$.
\begin{figure}[!htbp]\label{kap}
\center{\subfigure[~$ a=1, ~b=2 $] {
\includegraphics[angle=0,width=7cm,keepaspectratio]{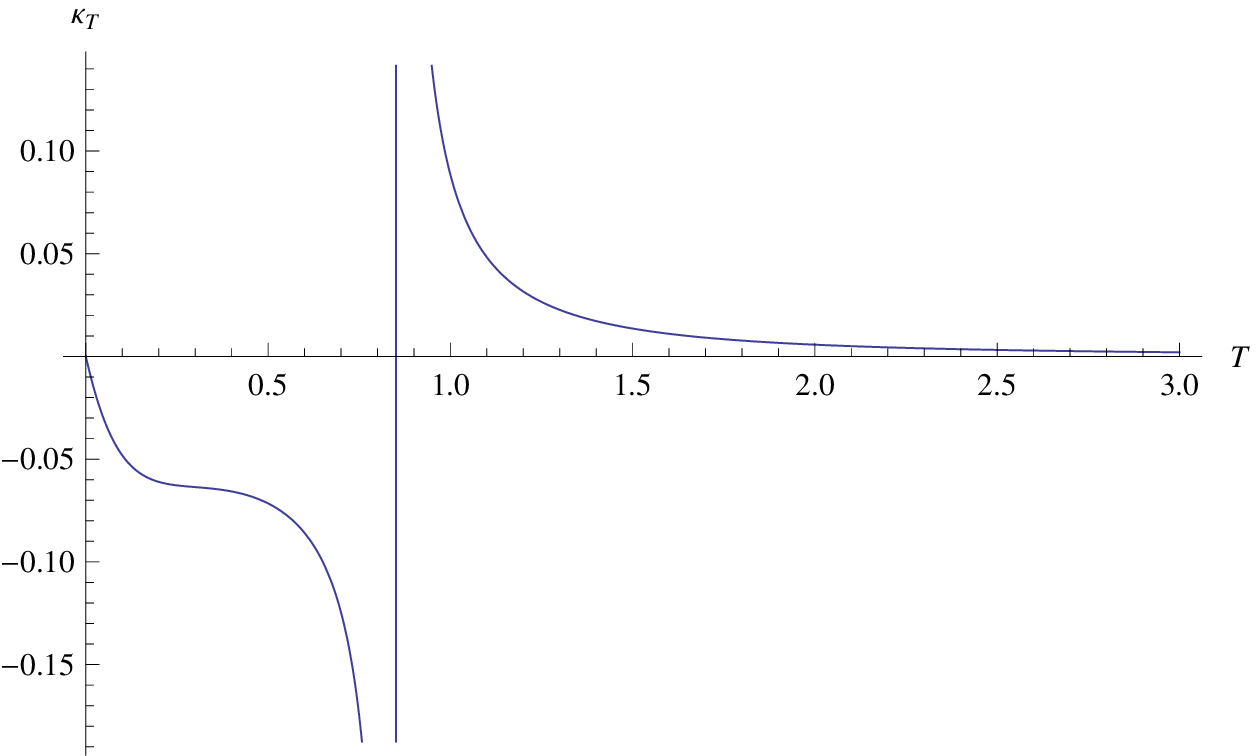}}\hfill
\subfigure[~$ a=-1, ~b=-2 $] {
\includegraphics[angle=0,width=7cm,keepaspectratio]{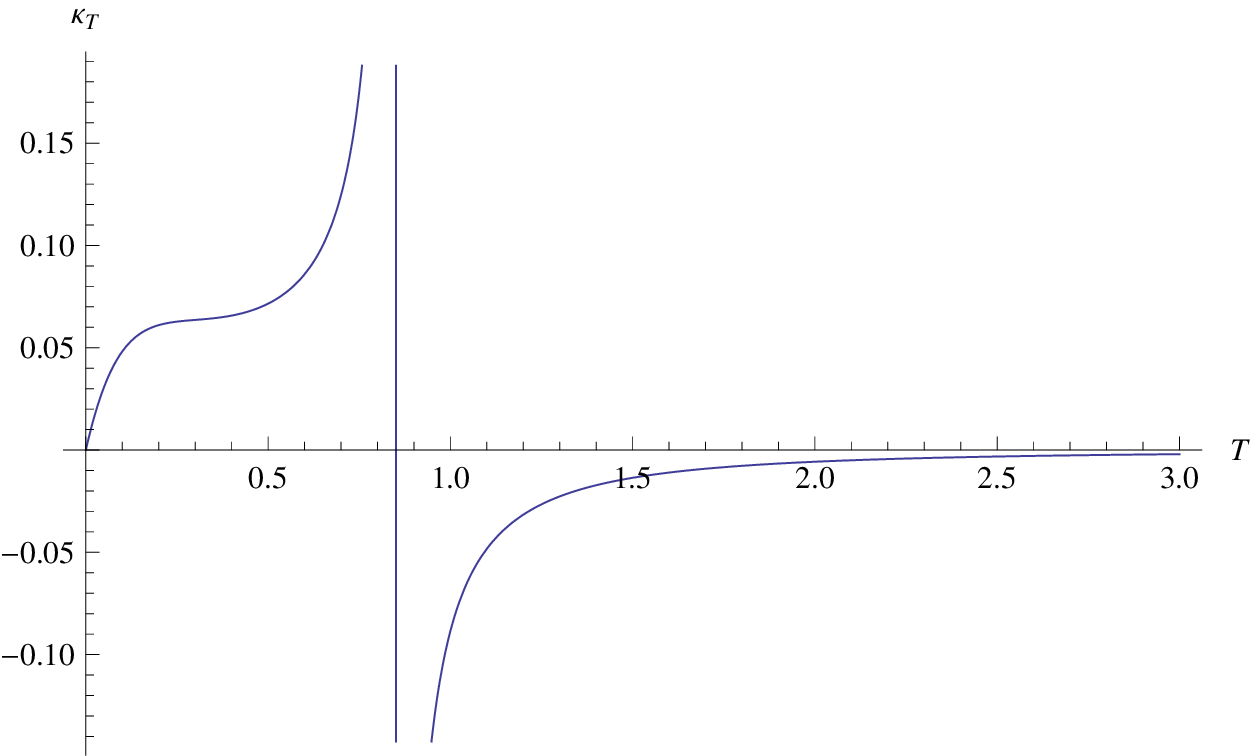}}
\caption[]{\it The isothermal compressibility  as functions of temperature for BTZT
 black hole for the choices of $ ~l=1, ~r_{-} =1,~a=\pm 1,~b=\pm 2$. }}
\end{figure}

In order to see the thermodynamic behavior near the critical point, the critical exponents can be introduced as
\ba\label{expo}
&&J-J_c\sim|\Omega-\Omega_c|^{\delta}, \quad
\Omega-\Omega_c\sim|T-T_c|^{\beta}\no\\
&&C_J\sim |T-T_c|^{-\alpha}, \quad
\kappa_T\sim |T-T_c|^{-\gamma}
\ea
From Eq.(\ref{JW}), at the critical point $T=T_c$, the first derivative of $J$ over $\Omega$ satisfy
\be
\left.{\frac{\partial J}{\partial \Omega}}\right|_{T_{c}} =0
\ee
But the second derivative can be calculated as
\be
\left.{\frac{\partial^2 J}{\partial \Omega^2}}\right|_{T_{c}}=\pm 322.187\neq 0
\ee
for the two cases $a=1,~b=2,~l=1$ and $a=-1,~b=-2,~l=1$. Thus
\be
J-J_c=\left.{\frac{\partial^2 J}{\partial \Omega^2}}\right|_{T_{c}}(\Omega-\Omega_c)^2+O((\Omega-\Omega_c)^3)
\ee
which means $\delta=2$.

According to Eq.(\ref{hc}),
\ba\label{CJ}
&&C_J=\no\\
&&-\frac{4 \pi  r_{+}^2 \left(b^2-a^2 l^2\right) \left(-2 a l r_{+} \left(\sqrt{a^2 l^2 r_{+}^2-b^2 r_{+}^2-b J l^2}+a l r_{+}\right)+2 b^2 r_{+}^2+b J l^2\right)}{l \left(2 a^3 l^3 r_{+}^3+2 a^2 l^2 r_{+}^2 \sqrt{a^2 l^2 r_{+}^2-b^2 r_{+}^2-b J l^2}+b \left(J l^2-2 b r_{+}^2\right) \sqrt{a^2 l^2 r_{+}^2-b^2 r_{+}^2-b J l^2}-2 a b^2 l r_{+}^3\right)}
\ea
One can set
\be\label{critical}
T=T_c(1+\epsilon), \quad r_+=r_c(1+\Delta)
\ee
where $|\epsilon|,~|\Delta|\ll1$.
Because the entropy $S$ and the temperature $T$ can both be expressed as $S=S(r_{+},J),~T=T(r_{+},~J)$, and
\be
C_J=T\left.{\frac{\partial S}{\partial T}}\right|_{J}=T\frac{\left.{\frac{\partial S}{\partial r_{+}}}\right|_{J}}
{\left.{\frac{\partial T}{\partial r_{+}}}\right|_{J}}
\ee
According to Fig.\ref{CPR}, at the critical point $r=r_c$
\be
\left.\left({\frac{\partial T}{\partial r_{+}}}\right)_{J}\right|_{r=r_c}=0
\ee
Moreover one can easily verify  $\left.\left({\frac{\partial^2 T}{\partial r_{+}^2}}\right)_{J}\right|_{r=r_c}\neq0$.
Therefore, in a sufficiently small neighborhood of $r_c$, one can expand $T$ in terms of $r_{+}$ as
\be
T(r_{+})=T(r_c)+\d{1}{2}\left.\left({\frac{\partial^2 T}{\partial r_{+}^2}}\right)_{J}\right|_{r=r_c}r_c^2\Delta^2+O(\Delta^3)
\ee
from which we obtain
\be
\Delta=\d{\epsilon^{1/2}}{D^{1/2}}
\ee
where
\be
D=\d{r_c^2}{2T_c}\left.\left({\frac{\partial^2 T}{\partial r_{+}^2}}\right)_{J}\right|_{r=r_c}
\ee
Substitute Eq.(\ref{critical}) into Eq.(\ref{CJ}), we can derive that the critical behavior of $C_J$ is described by
\be\label{alp}
C_J\approx\d{A}{\epsilon^{1/2}}
\ee
where $A$ is a function of $a,~b,~l,~J_c,~D$ and very complicated. Here we do not give the detailed expression. Comparing Eq.(\ref{alp}) with
Eq.(\ref{expo}), one can find that $\alpha=1/2$.

To calculate $\beta$ we first derive the $\Omega$ as function of $r_{+},~J$.
\be
\Omega(r_{+},J)=\frac{\sqrt{a^2 l^2 r_{+}^2-b^2 r_{+}^2-b J l^2}+a l r_{+}}{b l r_{+}}
\ee
For fixed $a,~b,~l$ and the critical $J_c$,
\be
\Omega(r_{+},J)=\Omega(r_{c},J_c)+\left.\left({\frac{\partial \Omega}{\partial r_{+}}}\right)_{J}\right|_{r=r_c}(r_{+}-r_{c})+\text{higher order terms}
\ee
Ignoring the higher order terms, we finally obtain
\be
\Omega(r_{+},J)-\Omega(r_{c},J_c)=\frac{J_c l}{r_{c}^2 \sqrt{a^2 l^2 r_{c}^2-b^2 r_{c}^2-b J_c l^2}T_c^{1/2}D^{1/2}}|T-T_c|^{1/2}
\ee
Therefore $\beta=1/2$.

Following the previous approach one can express the $\kappa_T$ as function of $r_{+},~J$. Utilizing Eq.(\ref{critical}) we can obtain
\be
\kappa_T\approx \d{B}{\Delta}=\d{BD^{1/2}}{\epsilon^{1/2}}=\d{BD^{1/2T_c^{1/2}}}{|T-T_c|^{1/2}}
\ee
which means $\gamma=1/2$.
Therefore the critical exponents $\alpha,~\beta,~\gamma,~\delta$ have the same values as the ones obtained in the Ho\v{r}ava-Lifshitz black hole and the Born-Infeld
black hole\cite{Majhi,Banerjee2}. Obviously they obey the scaling symmetry like the ordinary thermodynamic systems
\ba
&&\alpha+2\beta+\gamma=2, \quad \alpha+\beta(\delta+1)=2\no\\
&&\gamma(\delta+1)=(2-\alpha)(\delta-1), \quad \gamma=\beta(\delta-1)
\ea

\section{Discussion and conclusion}

In this paper, we adopted Ehrenfest's classification to study the phase transition of the BTZ black hole with torsion
obtained in the MB topological gravitational model. Although the
gravitational action contains torsion, the metric part of the BTZ black hole with torsion looks like the usual BTZ solution. Because
of the existence of  the Chern-Simons term and the Nieh-Yan term, the
conserved charges for the BTZ black hole should be modified.  Inclusion of these topological terms makes the thermodynamic properties and
critical behaviors of BTZ black hole with torsion very different from the ones of the usual BTZ black hole obtained in GR.

By treating the effective cosmological constant as a thermodynamic pressure, in the extended phase space we completely followed the standard of Ehrenfest to
explore the type of the phase transition of the BTZ black hole with torsion. It is shown that when $|a|l\leq|b|$ the Gibbs free energy and entropy are continuous
functions of temperature, however the heat capacity $C_P$, the isothermal compressibility $\kappa$ and the expansion coefficient $\alpha$
are all divergent at the critical point. This means this kind of phase transition for the BTZ black hole with torsion is continuous or second order. Nevertheless,
the phase transition and critical behavior are different from the ones in the van der Waals liquid/gas system.
Because $a,~b$ here are related to the parameters $\alpha_3,~\alpha_4$ in the action of MB model. Thus whether phase transition can happen depends not only upon
the black hole solutions, but also upon the gravitational actions.

Moreover we also considered the non-extended phase space. In this case, no direct thermodynamic
analogy for the the isothermal compressibility exists. Thus we employed another form and named it $\kappa_T$. It is shown that the $\kappa_T$
also diverge at the same critical point. Therefore in the non-extended phase space, the phase transition is also the second order.
The critical exponents for the BTZT black hole are also calculated, which are the same  as the ones obtained in the Ho\v{r}ava-Lifshitz black hole and the Born-Infeld
black hole. Is this just a coincidence, or is there some inherent reason, still need consideration further.

Although we discussed the three-dimensional topological model with torsion, the results have included the torsion-free case which corresponds to
the topologically massive gravity (TMG)\cite{DJT}. For the TMG, the field equations of which are also solved by the BTZ metric (CS-BTZ solution).
The conserved charges and the entropy are modified to be \cite{SN,Park,PKT}
 \be\label{charge}
 M=M_0-\frac{\beta}{
 L^2}J_0, \quad J=J_0-\beta M_0,\quad S=4\pi(r_+-\frac{\beta}{
L}r_{-})
\ee
Here $~\beta~$ is the Chern-Simons coupling constant and $L$ is the usual cosmological radius. Obviously the phase transition and the critical behaviors
 for the BTZ black hole
in the TMG correspond to the $a=1$ case  of the BTZT black hole in the MB model. Therefore when $|\beta|>L$ phase transition also exists in the CS-BTZ black hole.
Similarly, in this time the mass, angular momentum and the entropy cannot all be  positive.

\bigskip

\section*{Acknowledgements}
MSM thanks H. H. Zhao and H. F. Li for useful discussion. This work is supported in part by NSFC under Grant
Nos.(11247261;11175109;11075098;11205097).

\end{document}